\documentclass[apj,onecolumn]{emulateapj}

\shorttitle{The Doppler--21-cm correlation}
\shortauthors{Alvarez, Komatsu, Dor\'e and Shapiro}

\begin{document}
\title{%
  The cosmic reionization history as revealed by 
  the CMB Doppler--21-cm correlation
}%

\author{%
  Marcelo A. Alvarez\altaffilmark{1},
  Eiichiro Komatsu\altaffilmark{1},
  Olivier Dor\'e\altaffilmark{2},
  and Paul R. Shapiro\altaffilmark{1}
}%

\altaffiltext{1}{%
  Department of Astronomy, University of Texas at Austin,
  1 University Station, C1400, Austin, TX 78712
}%
\altaffiltext{2}{%
Canadian Institute for Theoretical Astrophysics, University of
  Toronto, 60 St. George Street, Toronto, ON M5S 3H8, Canada
}%

\begin{abstract}
  We show that the epoch(s) of reionization when the average
  ionization fraction of the universe is about half can be determined
  by correlating Cosmic Microwave Background (CMB) temperature maps
  with 21-cm line maps at degree scales ($l\sim 100$). During
  reionization peculiar motion  of free electrons induces the Doppler
  anisotropy of the CMB, while density fluctuations of neutral
  hydrogen induce the 21-cm line anisotropy.  In our simplified model
  of inhomogeneous reionization,
  a positive correlation arises as the universe reionizes whereas a
  negative correlation arises as the universe recombines; thus, the
  sign of the correlation provides information on the
  reionization history which cannot be obtained by present means.  
  The signal comes mainly from large scales
  ($k\sim 10^{-2}~{\rm Mpc}^{-1}$) where linear perturbation theory is
  still  valid and complexity due to patchy reionization is averaged
  out.  Since the Doppler signal comes from ionized regions and the
  21-cm comes from neutral ones, the correlation has a well defined peak(s)
  in redshift when the average ionization fraction of the universe
  is about half.  Furthermore, the cross-correlation is much less
  sensitive to systematic errors, especially foreground emission, than
  the auto-correlation of 21-cm lines:  this is analogous to the
  temperature-polarization correlation of the CMB being more immune to
  systematic errors than the  polarization-polarization. Therefore, we
  argue that the Doppler-21cm correlation provides a robust
  measurement of the 21-cm anisotropy, which can also be used as a
  diagnostic tool for  detected signals in the 21-cm data --- 
  detection of the cross-correlation provides the strongest confirmation 
  that the detected signal is of cosmological origin.
  We show that the Square Kilometer Array can easily measure the 
  predicted correlation signal for 1~year of survey observation.
\end{abstract}


\section{Introduction}\label{sec:introduction}

When and how was the universe reionized?  This question is deeply
connected to the physics of formation and evolution of the first
generations of ionizing sources (stars or quasars or both) and the
physical conditions in the interstellar and the intergalactic media in
a high redshift universe.  This field has been developed mostly
theoretically \citep{barkana/loeb:2001,bromm/larson:2004,ciardi/ferrara:2005,
iliev/etal:2005,alvarez/bromm/shapiro:2005} because there are only 
a very limited number of  observational probes of the epoch of
reionization: the Gunn--Peterson test
\citep{gunn/peterson:1965,becker/etal:2001}, polarization of the
Cosmic Microwave Background (CMB)  on large angular scales
\citep{zaldarriaga:1997,kaplinghat/etal:2003,kogut/etal:2003}, mean
intensity
\citep{santos/bromm/kamionkowski:2002,salvaterra/ferrara:2003,cooray/yoshida:2004,madau/silk:2005,fernandez/komatsu:2005} 
and fluctuations
\citep{magliocchetti/salvaterra/ferrara:2003,kashlinsky/etal:2004,cooray/etal:2004,kashlinsky/etal:2005}
of the near infrared background from redshifted UV photons, 
Ly$\alpha$-emitters at high redshift 
\citep{malhotra/rhoads:2004,santos:2004,furlanetto/hernquist/zaldarriaga:2004,haiman/cen:2005,wyithe/loeb:2005}
and fluctuations of the 21-cm line background from neutral hydrogen atoms
during reionization
\citep{ciardi/madau:2003,furlanetto/sokasian/hernquist:2004,zaldarriaga/furlanetto/hernquist:2004}
or even prior to reionization
\citep{scott/rees:1990,madau/meiksin/rees:1997,tozzi/etal:2000,iliev/etal:2002,shapiro/etal:2005}.

Each one of these methods probes different epochs and aspects of 
cosmic reionization: the Gunn--Peterson test is sensitive to a very small
amount of residual neutral hydrogen present at the late stages of
reionization ($z\sim 6$), Ly$\alpha$-emitting galaxies and
the wavelengths of the near infrared background probe the 
intermediate stages of reionization ($7\lesssim z \lesssim 15$), 
the 21-cm background probes the earlier stages where the majority of 
the intergalactic medium is still neutral ($10\lesssim z\lesssim 30$), and  
the CMB polarization measures the column density of 
free electrons integrated over a broader redshift range 
($z\lesssim 20$, say).
Since different datasets are complementary, one expects that 
cross-correlations
between them add more information than can be obtained by 
each dataset alone. For example, the information content in the CMB
and the 21-cm background cannot be exploited fully until the 
cross-correlation is studied: 
if we just extract the power spectrum from each dataset, we do not exhaust
the information content in the whole dataset because we are ignoring
the cross-correlation between the two.
The cross-correlation always reveals more 
information than can be obtained from the datasets individually 
unless the two are perfectly correlated (and Gaussian) or totally uncorrelated.

Motivated by these considerations, we study the cross-correlation
between the CMB temperature anisotropy and the 21-cm background
on large scales.
We show that the CMB anisotropy from the Doppler effect 
and the 21-cm line background can be 
{\it anti}-correlated or correlated at degree scales ($l\sim 100$), 
and both the amplitude and the sign of the correlation tell us how 
rapidly the universe reionized or recombined, and locations of 
the correlation (or the anti-correlation) peak(s) in redshift space 
tell us when reionization or recombination happened. 
This information is difficult to extract from either the CMB or
the 21-cm data alone.
Our work is different from recent work on a similar subject by
 \citet{salvaterra/etal:2005}.
While they studied a similar cross-correlation on very small scales
($\sim$ arc-minutes), we focus on much larger scales ($\sim$ degrees)
where matter fluctuations are still linear
and complexity due to patchy reionization is averaged out. 
\citet{cooray:2004} studied higher-order correlations
such as the bispectrum on arc-minute scales.
For our case, however, fluctuations are expected to follow nearly Gaussian
statistics on large scales, and thus one cannot obtain more information 
from higher-order statistics.  He also studied the cross-correlation
power spectrum of the CMB and projected 21-cm maps, and concluded that
the signal would be too small to be detectable owing to the
line-of-sight cancellation of the Doppler signal in the CMB. However, we
show that cancellation can be partially avoided by cross-correlating
the CMB map with 21-cm maps at different redshifts
(tomography). Prospects for the Square Kilometer Array (SKA) to
measure the cross-correlation signal on degree scales are shown to be
promising. 

Throughout the paper, we use $c=1$ and the following convention for the Fourier transformation:
\begin{equation}
f(\hat{\mathbf n},\eta) = 
\int \frac{d^3{\mathbf k}}{(2\pi)^3}f_{\mathbf k}e^{-i{\mathbf k}\cdot{\hat{\mathbf n}}(\eta_0-\eta)},
\end{equation}
where $\hat{\mathbf n}$ is the directional cosine along the line of sight 
pointing toward the celestial sphere, $\eta$ is the conformal
time, $\eta(z)=\int_0^t dt'/a(t')=\int_z^\infty dz'/H(z')$, and $\eta_0$ is the conformal
time at present. Note that
\begin{equation}
\eta_0-\eta(z) = \int_0^z \frac{dz'}{H(z')},
\end{equation}
which equals the comoving distance, $r(z)=\eta_0-\eta(z)$, in flat 
geometry (with $c=1$). Also, using Rayleigh's formula one obtains
\begin{equation}
f(\hat{\mathbf n},\eta) = 
4\pi\sum_{lm}(-i)^l
\int \frac{d^3{\mathbf k}}{(2\pi)^3}f_{\mathbf k}
j_l[k(\eta_0-\eta)]Y_{lm}(\hat{\mathbf{n}})Y_{lm}^*(\hat{\mathbf k}).
\end{equation}
The cosmological parameters are fixed at 
$\Omega_m=0.3$, $\Omega_b=0.046$, $\Omega_\Lambda=0.7$, $h=0.7$, and
$\sigma_8=0.85$, and we assume a scale invariant initial power spectrum for 
matter perturbations. 

This paper is organized as follows. 
In \S~\ref{sec:21cmdop} and \ref{sec:dop21cm} we derive the analytic formula for the 
Doppler--21-cm correlation power spectrum.
Equations~(\ref{exact}) or (\ref{eq:exact_both}) are the main result.
We then present a physical picture of the correlation and
describe properties of the correlation in detail.
We also discuss the validity of our assumptions and possible effects of
more realistic reionization scenarios.
In \S~\ref{sec:exp} we discuss detectability of the correlation
signal with SKA before concluding in \S~\ref{sec:discussion}.

\section{21-cm Fluctuations and CMB Doppler Anisotropy}\label{sec:21cmdop}

\subsection{21-cm Signal}

Following the notation of \citet{zaldarriaga/furlanetto/hernquist:2004},
we write the observed differential brightness temperature of the 21-cm
emission line at $\lambda = 21~{\rm cm}(1+z)$ in the direction of
$\hat{\mathbf n}$ as 
\begin{equation}
  T_{21}(\hat{\mathbf{n}},z)=T_0(z)\int_0^{\eta_0} d\eta'W[\eta(z)-\eta']\psi_{21}
  (\hat{\mathbf{n}},\eta'),
\end{equation}
where $W(\eta(z)-\eta')$ is a normalized ($\int_{-\infty}^{\infty} dx W(x)=1$) 
spectral response function of an instrument
which is centered at $\eta(z)-\eta'=0$, $T_0(z)$ is a normalization factor given by
\begin{equation}
   T_0(z)\simeq 23~{\rm mK}~\left(\frac{\Omega_bh^2}{0.02}\right)
   \left[\left(\frac{0.15}{\Omega_mh^2}\right)\left(\frac{1+z}{10}\right)\right]^{1/2},
\end{equation}
and
\begin{equation}
\psi_{21}(\hat{\mathbf{n}},\eta)\equiv x_H(\hat{\mathbf{n}},\eta)[1+\delta_b(\hat{\mathbf{n}},\eta)]
\left[1-\frac{T_{\rm cmb}(\eta)}{T_{\rm s}(\hat{\mathbf{n}},\eta)}\right]\rightarrow 
\left\{1-{\overline{x}_e}(\eta)[1+\delta_x(\hat{\mathbf{n}},\eta)]\right\}
\left\{1+\delta_b(\hat{\mathbf{n}},\eta)\right\},
\label{psi21}
\end{equation}
where $\delta_b$ is the baryon density contrast, 
\begin{equation}
\delta_x\equiv \frac{x_e-{\overline{x}_e}}{\overline{x}_e},
\end{equation} 
is the ionized fraction contrast, $x_H$ is the neutral fraction, and $x_e\equiv 1-x_H$ is the
ionized fraction.
Here, we have assumed that the spin temperature of neutral hydrogen, 
$T_{\rm s}$, is much larger than the CMB temperature, $T_{\rm cmb}$.
This assumption is valid soon after reionization begins 
\citep{ciardi/madau:2003}.

To simplify the calculation, we assume that the spectral resolution of
the instrument
is much smaller than the features of the target signal in redshift space.
This is always a very good approximation.
(For the effect of a relatively large bandwidth, see 
\citet{zaldarriaga/furlanetto/hernquist:2004}.)
Therefore, we set $W(x) = \delta^D(x)$ to obtain 
$T_{21}(\hat{\mathbf{n}},z)=T_0(z)\psi_{21}[\hat{\mathbf{n}},\eta(z)]$.
To leading order in $\delta_x$ and $\delta_b$, the spherical harmonic 
transform of $T_{21}(\hat{\mathbf{n}},z)$ is given by
\begin{equation}
  a_{lm}^{21}(z)=4\pi (-i)^l\int \frac{d^3k}{(2\pi)^3}
  [\overline{x}_H(z)(1+f\mu^2){{\delta}}_{b\bf{k}}-{\overline{x}_e}(z){{\delta}}_{x\bf{k}}]
  \alpha_l^{21}(k,z)Y_{lm}^*({\bf{k}}),
  \label{alm21}
\end{equation}
where $\alpha^{21}_l(k,z)$ is a transfer function for the 21-cm line,
\begin{equation}
\label{alpha21}
  \alpha^{21}_l(k,z)\equiv T_0(z)D(z)j_l[k(\eta_0-\eta)],
\end{equation}
$D(z)$ is the growth factor of linear perturbations, $\mu\equiv
\hat{\mathbf{k}}\cdot \hat{\mathbf{n}}$, and $f\equiv d\ln D/d\ln a$.
The factor $(1+f\mu^2)$ takes account of the enhancement of the
fluctuation amplitude due to the redshift-space distortion, the
so-called ``Kaiser effect'' (Kaiser 1987; see also Bharadwaj \& Ali
2004 and Barkana \& Loeb 2005).

\subsection{Doppler Signal}

The CMB temperature anisotropy from the Doppler effect is given by
\begin{equation}
  T_D({\bf\hat{n}})= -T_{\rm cmb}
  \int_0^{\eta_0} d\eta \dot{\tau}e^{-\tau}{\bf\hat{n}}\cdot {\bf v}_b({\bf\hat{n}},\eta)
= -T_{\rm cmb}
\int_0^{\eta_0} d\eta \dot{\tau}e^{-\tau}{\bf\hat{n}}\cdot \int \frac{d^3k}{(2\pi)^3}
{{\mathbf v}_{b{\mathbf k}}}(\eta)e^{-i{\bf k}\cdot {\bf\hat{n}}(\eta_0-\eta)},
\label{doppler}
\end{equation}
where $T_{\rm cmb}=2.725~K$ is the present-day CMB temperature,
$\dot{x}\equiv \partial x/\partial\eta$, 
$\tau(\eta)\equiv \sigma_T\int_0^{\eta} d\eta' n_e(\eta')$ is the Thomson scattering optical depth,
and ${{\mathbf v}_{b{\mathbf k}}}$ is the peculiar velocity of baryons.
In deriving the above formula, we have neglected the fluctuation of
ionized fraction, $\delta_x(\hat{\mathbf{n}},\eta)$, 
and electron number density, $\delta_e(\hat{\mathbf{n}},\eta)$,
since their
contributions to the cross-correlation would be higher order corrections
(the effect due to $\delta_e{\mathbf v}_b$ 
is called the Ostriker--Vishniac effect; e.g., Ostriker \& Vishniac
1986), and such a correction is negligible for linear fluctuations on the large
scales we consider here. Note that the negative sign ensures that 
we see a blueshift, $T_D(\hat{\mathbf n}) >0$, when 
baryons are moving toward us, $\hat{\mathbf n}\cdot {\mathbf v}_{b}<0$.
The peculiar velocity is related to the density contrast via the
continuity equation for baryons, 
${{\mathbf v}_{b{\mathbf k}}}=-i({\bf k}/{k^2})\delta_{b\bf k}\dot{D}$.
One obtains
\begin{equation}
  T_D({\bf\hat{n}})=T_{\rm cmb}\int_0^{\eta_0} d\eta \dot{D}\dot{\tau}e^{-\tau}\int
  \frac{d^3k}{(2\pi)^3}
  \frac{\delta_{b\bf k}}{k^2}\frac{\partial}{\partial\eta}e^{-i{{\bf k\cdot\hat{n}}(\eta_0-\eta)}}.
\end{equation}
The spherical harmonic transform of $T_{D}(\hat{\mathbf{n}},z)$ is then given by
\begin{equation}
  a^{D}_{lm} = 4\pi (-i)^l\int \frac{d^3k}{(2\pi)^3}{{\delta}}_{b\bf{k}}
  \alpha^D_l(k)Y_{lm}^*({\bf{k}}),
\label{almD}
\end{equation}
where $\alpha^D_l(k)$ is a transfer function for the Doppler effect,
\begin{equation}
  \alpha^D_l(k)\equiv \frac{T_{\rm cmb}}{k^2}\int_0^{\eta_0} d\eta
  \dot{D}\dot{\tau}e^{-\tau}
  \frac{\partial}{\partial\eta} j_l[k(\eta_0-\eta)].
\label{alphaD}
\end{equation}

\section{Doppler--21-cm Correlation}\label{sec:dop21cm}

\subsection{Generic Formula}

Given the spherical harmonic coefficients just derived for the 21-cm line 
(Eq.~[\ref{alm21}]) and the Doppler anisotropy (Eq.~[\ref{almD}]),
one can calculate the cross-correlation power spectrum, $C_l^{21-D}$,
exactly as
\begin{eqnarray}
  \nonumber
  & &C^{21-D}_l(z) \\
  \label{cl}
  &=&\langle a^{21}_{lm}(z)a^{D*}_{lm}\rangle 
  =\frac{2}{\pi}\int_0^\infty k^2dk\left[\overline{x}_H(z)
  (1+f\langle \mu^2\rangle) P_{\delta\delta}(k)-\overline{x}_e(z)P_{x\delta}(k)\right]
  \alpha_l^{21}(k,z)\alpha_l^{D}(k) \\
  \nonumber
  &=& T_{\rm cmb}T_0(z)D(z)\frac{2}{3\pi}
      \int_0^\infty dk \left[4\overline{x}_H(z)P_{\delta\delta}(k)-
      3\overline{x}_e(z)P_{x\delta}(k)\right]j_l[k(\eta_0-\eta)]
      \int_0^{\eta_0} d\eta'\dot{D}\dot{\tau}e^{-\tau}
      \frac{\partial}{\partial\eta'}j_l[k(\eta_0-\eta')],
\end{eqnarray}
where we have 
defined the matter power spectrum, $P_{\delta\delta}(k)$, 
as $\langle\delta_{\bf k}\delta^*_{{\bf k}'}\rangle
\equiv (2\pi)^3\delta({\bf k}-{\bf k}')P_{\delta\delta}(k)$, and the
cross-correlation power spectrum between ionized fraction and density
$P_{x\delta}(k)$, as $\langle\delta_{x\bf k}\delta^*_{{\bf k}'}\rangle
\equiv (2\pi)^3\delta({\bf k}-{\bf k}')P_{x\delta}(k)$.  In the last
line of equation (\ref{cl}), we have used $f\langle \mu^2\rangle=1/3$
for a matter-dominated universe.
Note that $\delta$ used in these power spectra is the density contrast 
of {\it total} matter, $\delta$, as baryons trace total matter perturbations,
$\delta_b=\delta$, on the scales of our interest 
(scales much larger than the Jeans length of baryons).
Equation (\ref{cl}) can be simplified by integrating it by parts:
\begin{eqnarray}
  \nonumber
  C^{21-D}_l(z)&=&-T_{\rm cmb}T_0(z)D(z)\frac{2}{3\pi}
  \int_0^{\eta_0} d\eta' \frac{\partial}{\partial\eta'}\dot{D}\dot{\tau}e^{-\tau}\\
  &\times &\int_0^\infty dk 
  \left[4\overline{x}_H(z)P_{\delta\delta}(k)-3\overline{x}_e(z)P_{x\delta}(k)\right]
  j_l[k(\eta_0-\eta)]j_l[k(\eta_0-\eta')].
  \label{exact}
\end{eqnarray}
A further simplification can be made by using an approximation to the
integral of the product of spherical Bessel functions for $l\gg 1$: 
\begin{equation}
  \frac{2}{\pi}\int_0^\infty dk P(k)j_l(kr)j_l(kr') \approx
  P\left(k=\frac{l}{r}\right)\frac{\delta(r-r')}{l^2},
  \label{limber}
\end{equation}
where $r(z) = \eta_0-\eta(z)$ is the comoving distance out to an object
at a given $z$.
We obtain
\begin{equation}
  l^2C_l^{21-D}(z)\approx -T_{\rm cmb}T_0(z)
  D(z)
  \left[\frac{4}{3}\overline{x}_H(z)P_{\delta\delta}\left(\frac{l}{r(z)}\right)-
  \overline{x}_e(z)P_{x\delta}\left(\frac{l}{r(z)}\right)\right]
  \frac{\partial}{\partial\eta}(\dot{D}\dot{\tau}e^{-\tau}).
\label{crossapprox}
\end{equation}
In what follows we will use the exact expression given by equation
(\ref{exact}) in our main quantitative results, while we will retain
the approximate expression given by equation (\ref{crossapprox}) to
develop a more intuitive understanding of the origin of the
cross-correlation.  We have found that the exact expression gives
results which are about 
10\% lower (at $l\sim 100$) than the approximate expression of
equation (\ref{crossapprox}) for the single reionization history we
will use in \S~3.4, while for the double 
reionization history the exact result is smaller by about 40\%.
This is because the line-of-sight integral in equation (\ref{cl})
acts to smooth out features in redshift, an effect which dissappears when the
delta function is used in the approximation.  Since the double
reionization model fluctuates much more strongly in redshift
than the single reionization model, the effect is more apparent for
double reionization.

Equation (\ref{crossapprox}) implies one important fact: the cross-correlation
vanishes if $\dot{D}\dot{\tau}e^{-\tau}$ is constant.
In other words, the amplitude of the signal directly depends on
how rapidly structure grows and reionization proceeds,
and the sign of the correlation depends on the direction of reionization
(whether the universe recombines or reionizes).
Moreover, the shape of $l^2C_l(z)$ directly traces the shape of the matter
power spectrum at $k=l/r(z)$.
It is well known that $P(k)$ has a broad peak at the scale of 
the horizon size at the epoch of matter-radiation equality, 
$k_{\rm eq}\simeq 0.011~{\rm Mpc}^{-1}~(\Omega_mh^2/0.15)$.
Since the conformal distance (which is the same as the comoving
angular diameter distance in flat geometry) is on the order of 
$10^4$~Mpc at high redshifts, the correlation power spectrum
will have a peak at degree scales, $l\sim 10^2$.

\begin{figure*}
\plotone{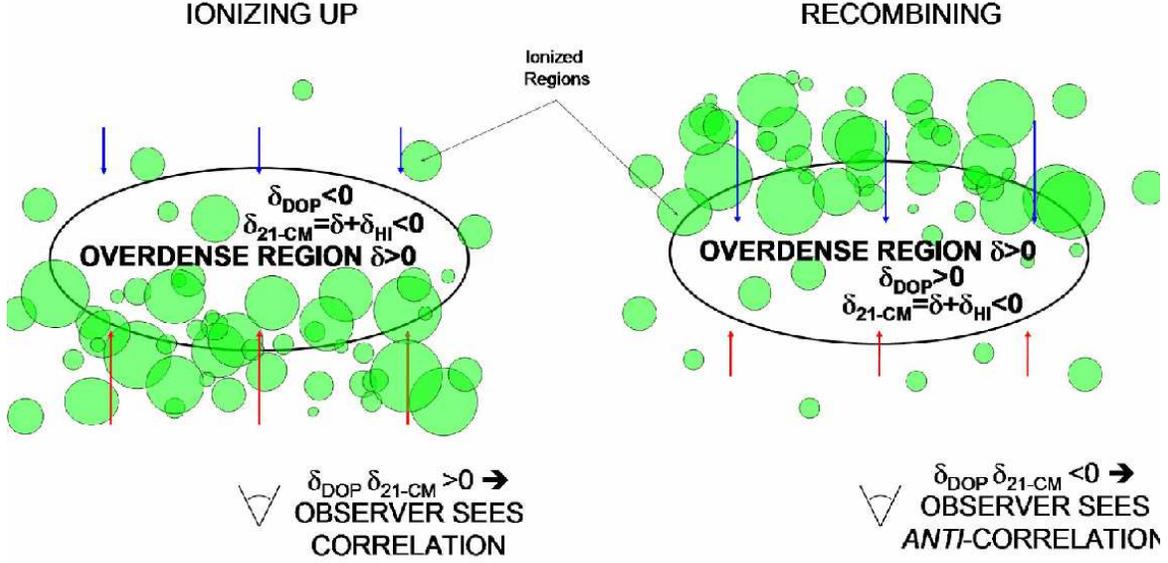}
\caption{Simplified schematic diagram illustrating the nature of the
correlation between the Doppler and 21-cm anisotropies.  Red arrows
pointing away from the observer indicate ionized gas falling into the positive
density perturbation (represented by the black oval) from the near
side, whereas blue arrows represent ionized gas falling in from the
far side.  During reionization, there is more ionized gas on the near
side of the perturbation (at lower redshift) than on the far side.
This implies that the net effect from this perturbation is a redshift
of the CMB in that direction (labeled as $\delta_{\rm DOP}<0$).
Because the sources responsible for reionization 
are located in halos which are very biased relative to the underlying
linear density field, the overdense region shown here is actually {\em
underdense} in neutral hydrogen, so that this overdensity represents a
negative fluctuation in the 21-cm signal (labeled as
$\delta_{\rm 21-cm}<0$).  Because both the 21-cm and 
the Doppler fluctuations from a region that is undergoing reionization
are both the same sign, the signature of reionization is a positive
correlation, while recombination (in which the situation is reversed
for the Doppler signal) results in an anti-correlation.
In reality, the growth of fluctuations and the
dependence of the density on redshift complicate the picture, so that
the sign of the signal is determined not by the derivative of the
ionized fraction $d[x_e]/dz$, but rather $d[x_e(z)(1+z)^{3/2}]/dz$ (see
equation \ref{deriv}).
\label{cartoon1}
}
\end{figure*}

\subsection{Ionized Fraction--Density Correlation}

While $P_{\delta\delta}(k)$ is a known function on the scales of
interest here, the cross-correlation between ionized fraction and
density, $P_{x\delta}(k)$, is not.  In order to understand its
importance in determining the observable signal, we have estimated its 
value on large scales.
Since we give the full details of derivations in Appendix, we quote only the 
result here:
\begin{equation}
\overline{x}_e(z)P_{x\delta}(k)= -\overline{x}_H(z)
\ln\overline{x}_H(z)\left[{\overline{b}}_h(z)-1-f\right]
P_{\delta\delta}(k),
\label{xdelta}
\end{equation}
where $\overline{b}_h(z)$ is the average bias of dark matter halos
more massive than $m_{\rm min}$,
\begin{equation}
{\overline{b}_h}(z)=1+\sqrt{\frac{2}{\pi}}\frac{e^{-\delta_c^2(z)/2
\sigma_{\rm min}^2}}{f_{\rm coll}(z)D(z)\sigma_{\rm min}},
\label{eq:p_xd}
\end{equation}  
$m_{\rm min}$ is the minimum halo mass capable of hosting ionizing
sources, $f_{\rm coll}(z)$ is the fraction of matter in the universe
collapsed into halos with $m>m_{\rm min}$, and $\sigma_{\rm min}\equiv
\sigma(m_{\rm min})$ is the r.m.s. of density fluctuations at the scale of
$m_{\rm min}$ at $z=0$.  
We take $m_{\rm min}$ to be the mass of a halo
with a virial temperature $T_{\rm min}$, which we will treat as a free
parameter. Here, $f$ is a parameter characterizing the physics of
reionization:  $f=0$ is the ``photon counting limit'', in which
recombinations are not important in determining the extent of ionized
regions.  On the other hand, $f=1$ is the ``Str\"omgren limit'', in
which the ionization rate is balanced by the recombination 
rate, as would occur in a Str\"omgren sphere.  While our choice for the range
of $f$ is resonable, larger values are possible if recombinations limit
the size of \ion{H}{2} regions and the clumping factor increases with
increasing density.  Equation (\ref{eq:p_xd}) is general in
the sense that it can accomodate such a scenario.  It is easy to check
that $P_{x\delta}$ naturally satisfies the physical constraints: it
vanishes when the universe is either fully neutral, $\overline{x}_H=1$, or
fully ionized, $\overline{x}_H=0$.
Although we have derived an explicit
relationship between $P_{x\delta}$ and $P_{\delta\delta}$ (which is
based on several simplifying assumptions -- see Appendix), we note
that the formulae presented here are sufficiently flexible so that any
model for the large-scale bias of reionization with
$P_{x\delta}=b_{x\delta}P_{\delta\delta}$ can be substituted for the
one we use here.

We simplify equations (\ref{cl}) and (\ref{crossapprox}) by making
the approximations that $e^{-\tau}\approx 1$ 
(justified by observations of the CMB polarization; see 
Kogut et al. 2003) and
\begin{equation}
D(z) = \frac{1+z_N}{1+z},
\end{equation}
which is a very good approximation at $z\gg 1$ when the universe 
is still matter-dominated.
The linear growth factor has been normalized such that $D(z_N)=1$; thus,
\begin{equation}
\dot{D}=
-H(z)\frac{d}{d z}\frac{1+z_N}{1+z}
=
\frac{(\Omega_m H_0^2)^{1/2}(1+z_N)}{(1+z)^{1/2}}.
\label{Ddot}
\end{equation}
We also use the relation
\begin{eqnarray}
\nonumber
  \dot{\tau}(z)&=&\sigma_T\frac{\rho_{b0}}{m_p}(1-Y_p)(1+z)^2{\overline{x}_e}(z)\\
  &=& 0.0525 H_0 \Omega_bh(1+z)^2 \overline{x}_e(z),
\label{taudot}
\end{eqnarray}
where  $\rho_{b0}$ is the baryon density at present, $Y_p=0.24$ is the
helium mass abundance (hydrogen ionization only is assumed), and ${\overline{x}_e}(z)$
is the ionized fraction.
One finds
\begin{equation}
\frac{\partial}{\partial\eta}(\dot{D}\dot{\tau}e^{-\tau})
=
-0.0525H_0^3
\Omega_m \Omega_b h
(1+z_{\rm N})(1+z)^{3/2}\frac{d}{dz}
\left[{\overline{x}_e}(z)(1+z)^{3/2}\right].
\label{deriv}
\end{equation}
Combining equations (\ref{cl}) and (\ref{deriv}), we obtain
\begin{eqnarray}
\nonumber
\frac{l^2C^{21-D}_l(z)}{2\pi}&=&0.37~\mu{\rm K}^2~
\left(\frac{\Omega_bh^2}{0.02}\right)^2
\left(\frac{\Omega_mh^2}{0.15}\right)^{1/2}
{\overline{x}}_H(z)
\left[4/3+\ln{\overline{x}}_H(z)\left({\overline{b}}_h-f-1\right)\right]\\
&\times&
\left(\frac{1+z}{10}\right)^{-1/2}
\int_0^\infty dz' \frac{F_l(z,z')}{H(z')}(1+z')^{3/2}
\frac{d}{dz'}\left[ {\overline{x}_e}(z')(1+z')^{3/2}\right],
\label{eq:exact_both}
\end{eqnarray}
where
\begin{equation}
F_l(z,z')=l^2\int_0^\infty dk \frac{P(k)(1+z_N)^2}{10^5 {\rm Mpc}^3}
j_l\left[k(\eta_0-\eta(z))\right]j_l\left[k(\eta_0-\eta(z'))\right].
\end{equation}
The approximation of equation (\ref{crossapprox}) and equation (\ref{deriv}) imply
\begin{eqnarray}
\nonumber
\frac{l^2C^{21-D}_l(z)}{2\pi}&\simeq &18.4~\mu{\rm K}^2~
\left(\frac{\Omega_bh^2}{0.02}\right)^2
\left(\frac{\Omega_mh^2}{0.15}\right)^{1/2}
{\overline{x}}_H(z)
\left[4/3+\ln{\overline{x}}_H(z)\left({\overline{b}}_h-f-1\right)\right]\\
&\times&
\frac{P_{\delta\delta}[l/r(z),z_N](1+z_N)^2}{10^5~{\rm Mpc}^3}
\frac{d}{dz}\left[{\overline{x}_e}(z)(1+z)^{3/2}\right]\left(\frac{1+z}{10}\right).
 \label{eq:approx_both}
\end{eqnarray}
Note that $P(k,z_N)(1+z_N)^2$ is independent of the normalization epoch,
$z_N$, for $z_N\gg 1$; the result is independent of the choice 
of $z_N$, as expected.
These equations are the main result of this paper, and we shall use these
results to investigate the properties of the correlation in more detail. 
Since we have a product of $d\overline{x}_e/dz$ and $\overline{x}_H$,
we expect that the largest contribution comes from the ``epoch of
reionization'' when $\overline{x}_e(z)$ changes most rapidly.
Therefore, by detecting the
Doppler--21-cm correlation peak(s), one can determine the epoch(s) of
reionization.  The sign of the cross-correlation is also very important.
The sign of the cross-correlation is determined by the sign of
the derivative term and the difference between
${\overline{x}}_HP_{\delta\delta}$ and ${\overline{x}}_eP_{x\delta}$.
For the case in which ${\overline{x}}_HP_{\delta\delta} > {\overline{x}}_eP_{x\delta}$,
the Doppler effect and the 21-cm emission are {\it anti}-correlated
when the ionized fraction, $\overline{x}_e(z)$, increases toward low $z$
faster than $(1+z)^{-3/2}$.  
For our simplified model of inhomogeneous
reionization (see Appendix), we find that
${\overline{x}}_HP_{\delta\delta} < {\overline{x}}_eP_{x\delta}$, however, 
and in this case we find a positive correlation as the universe is being 
reionized.  This is unique information that cannot be
obtained by present means. See Fig. \ref{cartoon1} for a schematic
diagram which describes the nature of the cross-correlation.  

\subsection{Illustration: Homogeneous Reionization Limit}

It may be instructive to study the nature of the signal by taking
the ``homogeneous reionization limit'', in which the ionized fraction
is uniform, $\delta_x\equiv 0$.  Such a situation may be more relevant
than our model for biased reionization if, for example, the photons
responsible for reionization have a very long mean free path or
clumping in denser regions cancels the effect of bias.
In the homogeneous limit, the approximate
formula (eq. \ref{eq:approx_both}) implies
\begin{eqnarray}
\nonumber
\frac{l^2C^{21-D}_l(z)}{2\pi}&\simeq &24.5~\mu{\rm K}^2~
\left(\frac{\Omega_bh^2}{0.02}\right)^2
\left(\frac{\Omega_mh^2}{0.15}\right)^{1/2}
\frac{P_{\delta\delta}[l/r(z),z_N](1+z_N)^2}{10^5~{\rm Mpc}^3}\\
&\times&
{\overline{x}}_H(z)
\frac{d}{dz}\left[{\overline{x}_e}(z)(1+z)^{3/2}\right]
\left(\frac{1+z}{10}\right).
 \label{eq:approx}
\end{eqnarray}
\begin{figure}
\plotone{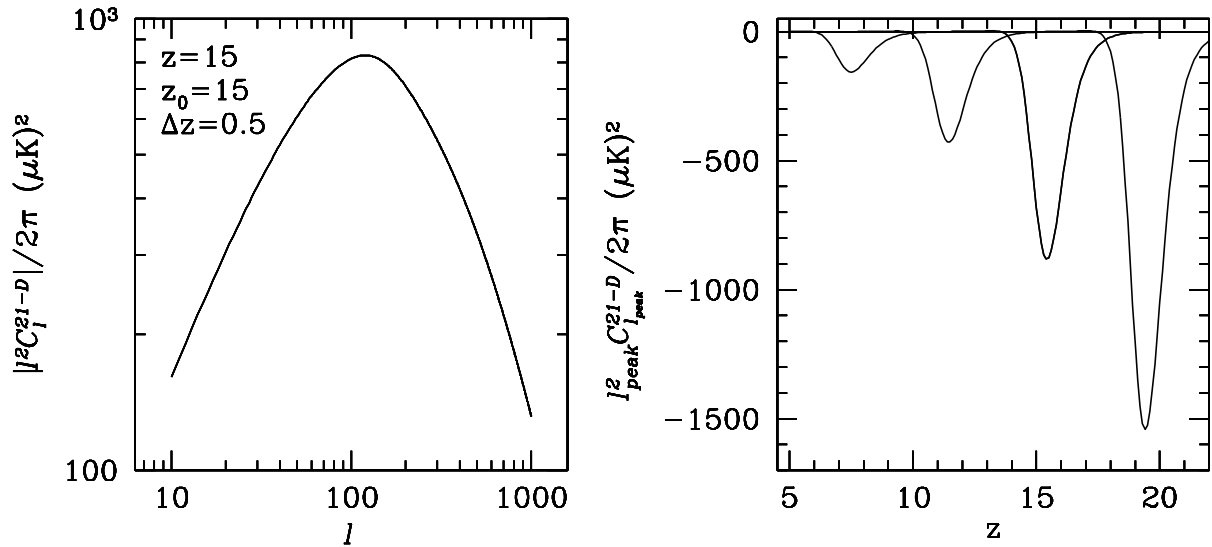}
\caption{({\it Left}) Power spectrum of the cross-correlation 
between the cosmic microwave background anisotropy and the 21-cm line
fluctuations, $l^2C_l^{21-D}/(2\pi)$.
We assume $z=15$ for a reionization history given by 
equation (\ref{zfh_param}) with a reionization redshift of
$z_r=15$ and duration of $\Delta z=0.5$.  
Note that its shape follows that of the linear 
matter power spectrum, $P_{\delta\delta}(k)$ with $k\simeq l/r(z)$, where 
$r(z)$ is the comoving angular diameter distance.  
({\it Right}) Evolution of the peak amplitude of $l^2C_l^{21-D}/(2\pi)$
at $l\sim 100$ from a homogeneous reionization history described by equation 
(\ref{zfh_param}), for $\Delta z=0.5$ and different reionization 
redshifts, $z_r=7,11,15$ and 19, from left to right.
\vspace{-0.5cm}
\label{plot1}
}
\end{figure}
One may estimate the amplitude of the signal
at the epoch of reionization, $z=z_r$, using a duration of reionization
at $z_r$, $\Delta z$ as follows (omitting factors of order unity):
\begin{equation}
\nonumber
\frac{l^2C^{21-D}_l(z_r)}{2\pi}\simeq -\frac{195~\mu{\rm K}^2}{\Delta z}
\frac{\overline{x}_H(z_r){\overline{x}_e}(z_r)}{0.25}
\left(\frac{1+z_r}{10}\right)^{5/2}.
\label{eq:order}
\end{equation}
The remarkable feature is that the predicted signal is rather large. 
For $z_r=15$ (which is consistent with early reionization suggested
by \citet{kogut/etal:2003}) and $\Delta z=1$, we predict 
$l^2C^{21-D}_l/(2\pi)\sim -600~\mu{\rm K^2}$ at $l\sim 10^2$.
Under these assumptions, therefore, detection of the anti-correlation
peak should not be too difficult, given that the Wilkinson
Microwave Anisotropy Probe ({\sl WMAP}) has already obtained an
accurate CMB temperature map at $l\sim 10^2$ \citep{wmap}.
When any experiment
for measuring the 21-cm background at degree scales becomes
on-line, one should correlate the 21-cm data on degree scales with the
{\sl WMAP} temperature map to search for this peak.  Note also that in
the homogeneous reionization limit the sign is reversed, so that
reionization results in an anti-correlation.  The sign of the
correlation therefore depends sensitively on the degree to which
reionization is biased on large scales.

\subsection{Reionization History}

To calculate the actual cross-correlation power spectrum, we need to
specify the evolution of the ionized fraction, $\overline{x}_e(z)$.
Computing $\overline{x}_e(z)$ from first principles is admittedly
very difficult, and this is one of the most challenging tasks in 
cosmology today. To illustrate how the cross-correlation power
spectrum changes for different reionization scenarios, therefore, we
explore two simple parameterizations of the reionization history. 

In one case, we assume that the ionized fraction increases monotonically
toward low $z$. We use the simple parameterization adopted by
\citet{zaldarriaga/furlanetto/hernquist:2004}:  
\begin{equation}
\overline{x}_H(z)=\frac{1}{1+{\rm exp}\left[-(z-z_r)/\Delta z\right]},
\label{zfh_param}
\end{equation}
where $z_r$ is the ``epoch of reionization'' when $\overline{x}_H(z_r)=1/2$
and $\Delta z$ corresponds to its duration.
In this case, one obtains a fully analytic formula for
the correlation power spectrum:
\begin{eqnarray}
\label{clapprox}
\nonumber
\frac{l^2C^{21-D}_l(z)}{2\pi}
&\simeq&58~\mu{\rm K}^2~
\left[4/3+\ln{\overline{x}}_H(z)\left({\overline{b}}_h-f-1\right)\right]
\frac{P[l/r(z),z_N](1+z_N)^2}{10^5~{\rm
Mpc}^3} 
\left(\frac{\Omega_bh^2}{0.02}\right)^2
\left(\frac{\Omega_mh^2}{0.15}\right)^{1/2}\\
&&\times
{\overline{x}}_H(z){\overline{x}}_e(z)\left[\frac{3}{2}-\frac{\overline{x}_H(z)(1+z)}{\Delta z}\right]
\left(\frac{1+z}{10}\right)^{3/2}.
\end{eqnarray}
In the homogeneous reionization limit, $P_{x\delta}\equiv 0$, 
one gets $l^2C_l/(2\pi)\simeq -165~\mu{\rm K}^2$ for 
$z=9=z_r$ and $\Delta z=1/2$, and
the amplitude of the signal scales as $(1+z_r)^{5/2}$, as expected
(see Eq.~[\ref{eq:order}]). 
For an early reionization at $z_r=15$,
the homogeneous reionization model predicts
$l^2C_l/(2\pi)\simeq -570~\mu{\rm K}^2$.

\begin{figure}
\plotone{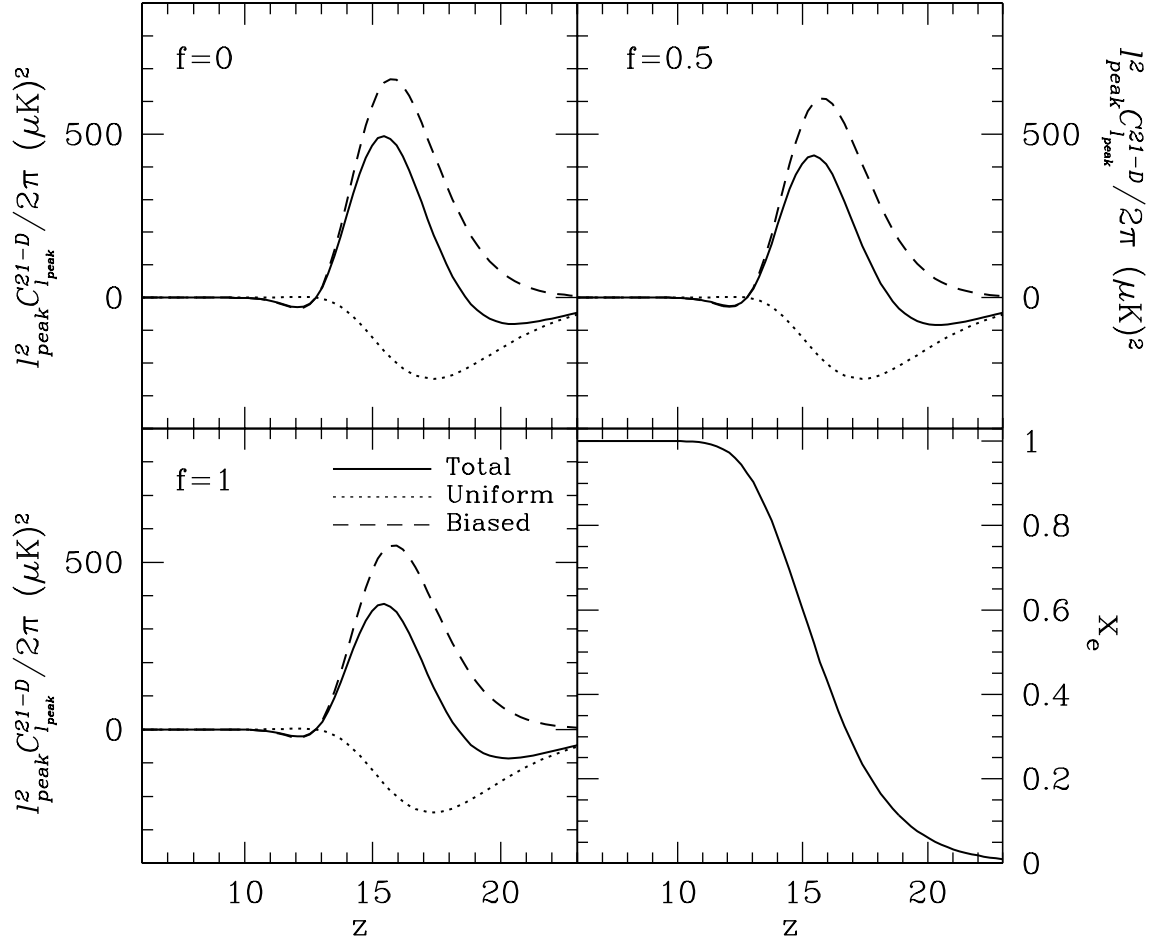}
\caption{
({\em top and left panels}) Peak correlation amplitude vs. redshift.  Each
panel is labeled with a different value of $f$ which parameterizes
the uncertainty in the physics of reionization
(see Appendix for the definition of $f$ and detailed discussion).  
The most likely value of $f$ is somewhere between 0 and 1.
The dotted line corresponds to the homogeneous reionization limit
in which fluctuations in the ionized fraction are totally ignored
(Eq.~[\ref{eq:approx}]), while the thick line takes into
account fluctuations in the ionized fraction
(Eq.~[\ref{eq:approx_both}]).
The dashed line is the difference between 
the homogeneous reionization and the total signal.
({\em bottom right}) Evolution of $\overline{x}_e$ with redshift.  Note that in all cases 
the reionization of the universe results in a positive correlation.
\vspace{-0.5cm}
} 
\label{plot2}
\end{figure}
The left panel of Figure~\ref{plot1} shows
the absolute value of the predicted correlation power spectrum,
$l^2C^{21-D}_l/(2\pi)$, for the homogeneous reionization model with
$z=15=z_r$ and $\Delta z = 0.5$.  
As we have explained previously, the shape of $l^2|C_l^{21-D}|$
 exactly traces that of the underlying linear matter power spectrum,
$P_{\delta\delta}$.
The right panel of Figure~\ref{plot1} shows the the redshift evolution of 
the peak value of the power spectrum at $l\sim 100$, 
for different values of $z_r$.
As discussed at the end of \S~3.2, the reionization of the universe 
leads to an anti-correlation between the Doppler and 21-cm fluctuations.  
The magnitude of the signal increases with redshift when 
the duration of reionization in redshift, $\Delta z$, is fixed
(see equation~(\ref{zfh_param})).
We could instead fix the duration of reionization in time, 
$\Delta t$, in which case $\Delta z$ increases with redshift as
$\Delta z \propto (1+z)^{5/2}\Delta t$; according to equation 
(\ref{eq:order}), 
therefore, the peak height in this case would be approximately independent 
of redshift.

To gain more insight into how the prediction changes with 
the details of the reionization process, let us use a somewhat
more physically motivated model for the ionized fraction,
\begin{equation}
\ln[1-{\overline{x}}_e(z)]=-\zeta_0(z)f_{\rm coll}(z).
\label{eq:xe}
\end{equation}
The ionized fraction increases monotonically toward low $z$ when
$\zeta_0$ does not depend on $z$.
Using this model with $\zeta_0=200$ and $T_{\rm min}=10^4$ K, 
we calculate the cross-correlation power spectrum.
Figure~\ref{plot2} plots the peak value of $l^2C_l^{21-D}$ as a function
of $z$, showing the contribution from $P_{\delta\delta}$, 
$P_{x\delta}$, and the sum of the two (Eq.~[\ref{eq:approx_both}]).
The bottom-right panel shows the evolution of the ionized
fraction predicted by equation~(\ref{eq:xe}).
In this figure we explore the dependence of the signal on the details 
of reionization by varying the parameter $f$. 
(See Appendix for the precise meaning of $f$.)
In all cases, the contribution from $P_{\delta\delta}$ is negative, whereas
that from $P_{x\delta}$ is positive; because the halo bias is
relatively large for our fiducial case of $T_{\rm min}=10^4$ K, with
$4<{\overline{b}}_h<17$ for $10<z<30$, the $P_{x\delta}$ term
dominates over the $P_{\delta\delta}$ term, and the correlation is
positive (see also Iliev et al. 2005). 
Increasing the value of $f$ towards a more recombination
dominated scenario decreases the importance of the dominant
$P_{x\delta}$ term, reducing the total amplitude of the signal further.
\begin{figure}
\plotone{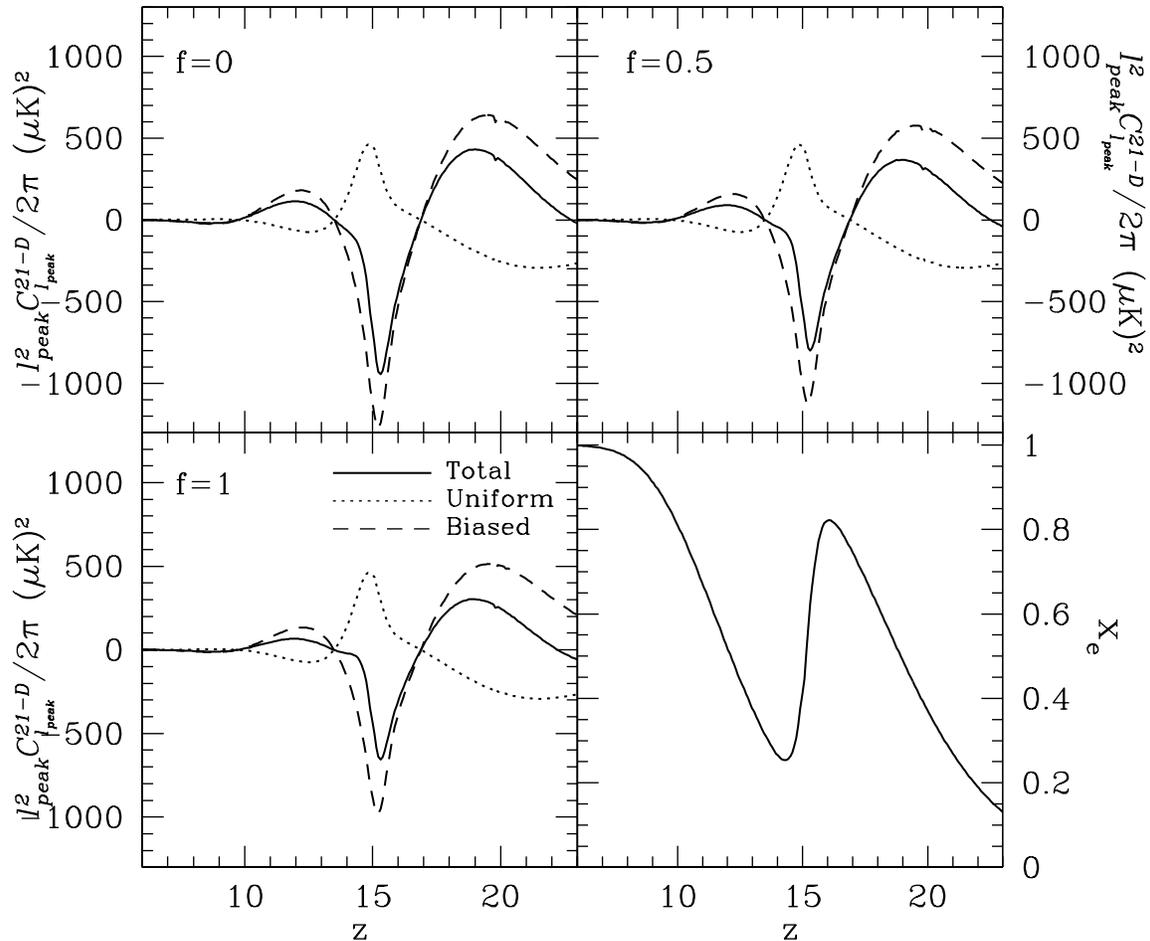}
\caption{
Same as in Fig. 2, but for a ``double reionization''
model in which the universe undergoes a brief period of recombination.  
Note that in all cases the recombination epoch results in a negative
correlation.
\vspace{-1.cm}
} 
\smallskip
\label{plot3}
\end{figure}
What happens when the universe was reionized twice (Cen 2003; see
however Furlanetto \& Loeb 2005)?
In Figure~\ref{plot3} we showed the case where the ionized fraction is
a monotonic function of redshift.  As seen in the figure, there is a
prominent correlation peak, regardless of the details of reionization 
process, encoded in $f$.  The situation changes completely when the
universe was reionized twice. We parameterize such a double
reionization scenario using a $z$-dependence for $T_{\rm min}(z)$ and
$\zeta_0(z)$: 
\begin{equation}
\zeta_0(z)=\zeta_i+(\zeta_f-\zeta_i)g(z)
\end{equation}
and
\begin{equation}
T_{\rm min}(z)=T_i+(T_f-T_i)g(z),
\end{equation}
where
\begin{equation}
g(z)=\frac{{\rm exp}\left[-(z_{\rm crit}-z)/\Delta z_{\rm tran}\right]}
{1+{\rm exp}\left[-(z_{\rm crit}-z)/\Delta z_{\rm tran}\right]}
\end{equation}
is a function that approaches zero for $z>z_{\rm crit}$ and unity for 
$z<z_{\rm crit}$, with a transition of duration $\Delta z_{\rm tran}$.
We take $z_{\rm crit}=15$, $\Delta z_{\rm tran}=0.25$, $\zeta_i=100$,
$\zeta_f=40$, $T_i=10^3$ K, and $T_f=10^4$ K.  In this case, the
minimum source halo virial temperature makes a smooth transition from
$10^3$ K at high redshift to $10^4$ K at low redshift, as might occur
if dissociating radiation suppresses star formation in ``minihalos''
with virial temperatures $<10^4$ K. The drop in $\zeta_0(z)$, which
for convenience coincides with the transition in $T_{\rm min}$,
could be due to, for example, metal pollution from Pop III stars
creating a transition to Pop II, accompanied by a transition from a
very top heavy IMF to a less top heavy one (e.g. Haiman \& Holder
2003).  In this scenario, the universe may recombine until enough Pop
II stars and halos with virial temperatures $>10^4 K$ form to finish
reionization. We emphasize that this is a simple parameterization for
illustration, and is not meant to represent a realistic double
reionization model. However, this model is sufficient to show that the
signature of a recombination epoch during reionization is a reversal
in the sign of the correlation.  Because of a rapid change in the
ionized fraction during recombination, the negative correlation peak is very
prominent, reaching $l^2C_l^{21-D}\sim 700-900~{\rm \mu K}^2$ for $f=0-1$. 

\section{Prospects for Detection}
\label{sec:exp}

\subsection{Error Estimation}

Assuming that CMB and instrumental noise for 21-cm lines are Gaussian, 
one can estimate the error of the correlation power spectrum by
\begin{equation}
(\Delta C_l)^2
= 
\frac1{(2l+1)f_{\rm sky}\Delta l}
\left[C_l^{\rm cmb}C_l^{\rm 21}+ (C_l^{\rm 21-D})^2\right],
\label{eq:error}
\end{equation}
where $\Delta l$ is the size of bins within which the power spectrum
data are averaged over $l-\Delta l/2<l<l+\Delta l/2$, 
and $f_{\rm sky}$ is a fraction of sky covered by observations,
\begin{equation}
f_{\rm sky} \equiv \frac{\Omega}{4\pi} = 2.424\times 10^{-3}
\left(\frac{\Omega}{100~{\rm deg^2}}\right).
\end{equation}
In the $l$ range we are considering ($l\sim 10^2$), CMB is totally
dominated by signal (i.e., noise is negligible), which gives
$l^2C_l^{\rm cmb}/(2\pi)\sim (50~\mu{\rm K})^2$ at $l\sim 10^2$.
On the other hand, the 21-cm lines will most likely be totally dominated 
by noise and/or foreground and the intrinsic signal contribution to the 
error may be ignored. We also assume that the foreground cleaning reduces
it to below the noise level.
We calculate the noise power spectrum based upon equation~(59) of
\citet{zaldarriaga/furlanetto/hernquist:2004}:
\begin{equation}
\frac{l^2 C_l^{\rm 21}}{2\pi}
=
\frac1{\Delta\nu t_{\rm obs}}
\left(\frac{ll_{\rm max}}{2\pi}\frac{\lambda^2}{A/T}\right)^2,
\label{eq:uniformuv}
\end{equation}
where $\Delta \nu$ is the bandwidth, $t_{\rm obs}$ is the total integration 
time, 
and $A/T$ is ``sensitivity'' (an effective area divided by system temperature)
measured in units of ${\rm m^2~K^{-1}}$.
The maximum multipole, $l_{\rm max}$, for a given baseline length, $D$, 
is given by
\begin{equation}
l_{\rm max}= 2\pi \frac{D}{\lambda}
= 2994\left(\frac{D}{\rm 1~km}\right)\left(\frac{10}{1+z}\right).
\end{equation}
Here, we have used $\lambda = 21~{\rm cm}(1+z)$.  Note that we have
implicitly assumed uniform coverage for the interferometer in deriving
equation~(\ref{eq:uniformuv}), which may not be realistic. Making the
baseline distribution more compact would enhance the detectability.

\subsection{Square Kilometer Array}

The current design of the Square Kilometer Array (SKA) aims 
at a sensitivity of $A/T \sim 5000~{\rm m^2~K^{-1}}$ at 
200~MHz.~\footnote{Information on SKA is available at {\sf 
http://www.skatelescope.org/pages/concept.htm}.}
Of which, 20\% of the area forms a compact array configuration
within a 1~km diameter, whereas 50\% is within 5~km, and 75\% is 
within 150~km.
Since we are interested in a relatively low-$l$ part of the spectrum,
we use the compact configuration, $D=1$~km, and 
$A/T = 5000\times 0.2=1000~{\rm m^2~K^{-1}}$.
We obtain
\begin{equation}
\frac{l^2 C_l^{\rm 21}}{2\pi}
=
\frac{(130~\mu{\rm K})^2}{N_{\rm month}\Delta\nu_{\rm MHz}}
\left[
\left(\frac{l}{100}\right)
\left(\frac{1+z}{10}\right)
\left(\frac{D}{\rm 1~km}\right)
\left(\frac{\rm 10^3~m^2~K^{-1}}{A/T}\right)
\right]^2,
\end{equation}
where $N_{\rm month}$ is the number of months of observations
and $\Delta\nu_{\rm MHz}$ is the bandwidth in units of MHz.
Note that we have assumed here that all the time spent during the
observation is on-source integration time.  However, a more realistic
assesment would be that a smaller (e.g., $\sim 1/3$) fraction of the
total time is spent integrating.  In this case, one should compensate
by increasing the total time of the observation accordingly.
Since the noise power spectrum is much larger than the amplitude of 
the predicted correlation signal, we safely ignore the contribution
of $C_l^{21-D}$ to the error. 
(We ignore the second term on the right hand side of Eq.~[\ref{eq:error}]).

The planned contiguous imaging field of view of SKA
is currently 1~${\rm deg^2}$ at $\lambda=21$~cm 
and it scales as $\lambda^2$. Using the number of independent survey
fields, $N_{\rm field}$, the total solid angle covered by observations
is given by $\Omega \simeq 100~{\rm deg^2}[(1+z)/10]^2 N_{\rm field}.$
This estimate is, however, based on the current specification for the
high frequency observations, and may not be relavant to low frequency
observations that we discuss here. It is likely that there will be
different telescopes with much larger field of view at low frequencies,
and thus we shall adopt a field of view which is three times larger:
\begin{equation}
\Omega \simeq 300~{\rm deg^2}\left(\frac{1+z}{10}\right)^2 N_{\rm field}.
\end{equation}
Using equation~(\ref{eq:error}) and the parameters of SKA, we find
the expected error per 
$\sqrt{N_{\rm month}N_{\rm field}\Delta\nu_{{\rm MHz}}}$ to be on the order of
\begin{equation}
{\rm Err}\left(\frac{l^2 C_l}{2\pi}\right)
\simeq
938~\mu{\rm K}^2
\sqrt{\frac{l/\Delta l}{N_{\rm month}N_{\rm field}\Delta\nu_{\rm MHz}}
\frac{l^2C_l^{\rm cmb}/(2\pi)}{2500~\mu{\rm K}^2}}.
\end{equation}
Therefore, for the nominal survey parameters,
$N_{\rm month}=12$ and $N_{\rm field}=4$, the SKA sensitivity 
to the cross-correlation power spectrum reaches
${\rm Err}[l^2C_l^{21-D}/(2\pi)]\simeq 135~{\rm \mu K}^2$,
which gives $\sim 3$-$\sigma$ detection of the correlation
peak for the normal reionization model, and 
$\sim 6$-$\sigma$ detection of the anti-correlation
peak for the double reionization model.
Increasing the integration time or the survey fields will obviously
increase the signal-to-noise ratio as 
$\sqrt{N_{\rm month}N_{\rm field}}$.
One would obtain more signal-to-noise by choosing a larger value for
$\Delta\nu$, which is equivalent to stacking different frequencies.
(However, $\Delta\nu$ must not exceed the width of the signal in
frequency space.)
Therefore, we conclude that  the cross-correlation
between the Doppler and 21-cm fluctuations is fairly easy
with the current SKA design. For more accurate measurements
of the shape of the spectrum, however, a larger contiguous
imaging field of view may be required. 
The more promising way to reduce errors may be to
increase sensitivity (i.e., larger $A/T$)
by having more area, $A$, for the compact configuration.
This is probably the most economical way to improve the signal-to-noise
ratio, as the error is linearly proportional to $(A/T)^{-1}$,
rather than the square-root.


\section{Discussion and Conclusion}\label{sec:discussion}

We have studied the cross-correlation between the CMB temperature
anisotropy and the 21-cm background.  
The cross-correlation occurs via the peculiar velocity field
of ionized baryons, which gives the Doppler anisotropy in CMB,
coupled to density fluctuations of neutral hydrogen, which  
cause 21-cm line fluctuations.
Since we are concerned with anisotropies in the cross-correlation on 
degree angular
scales ($l\sim 100$), which correspond to hundreds of comoving Mpc at
$z\sim 10$, we are able to treat density and velocity
fluctuations in the linear regime.  This greatly simplifies the
analysis, and distinguishes our work from previous work
on similar subjects that dealt only with the cross-correlation
on very small scales \citep{cooray:2004,salvaterra/etal:2005}.
Furthermore, because the 21-cm signal contains redshift
information, the cross-correlation is not susceptible to the line of
sight cancellation that is typically associated with the Doppler effect.
Finally, because the systematic errors of the 21-cm and CMB
observations are uncorrelated, the cross-correlation will be immune to
many of the pitfalls associated with observing the high redshift
universe in 21-cm emission, such as contamination by 
foregrounds\footnote{
A potential source of foreground contamination is the Galactic
synchrotron emission affecting both the CMB and 21-cm
fluctuation maps; however, the amplitude of synchrotron
emission in the CMB map at degree scales is much smaller
than the Doppler anisotropy from reionization, and thus
it is not likely to be a significant source of contamination.
}. 
We argue that detection of the predicted cross-correlation signal
provides the strongest confirmation that the signal detected
in the 21-cm data is of cosmological origin.
Without using the cross-correlation, it would be quite challenging 
to convincingly show that the detected signal does not come from
other contaminating sources.

We find that the evolution of the cross-correlation with redshift can
constrain the history of reionization in a distinctive way. In
particular, we predict that 
a universe undergoing reionization results in a positive
cross-correlation at those redshifts, whereas a recombining universe
results in a negative correlation (this dependends on our
simplified model of biased reionization -- a model in which
reionization is homogeneous would imply a reversal of the sign of
the correlation).
Thus, the correlation promises to
reveal whether the universe underwent a period of recombination during
the reionization process (e.g., Cen 2003), and to reveal the nature of
the sources of ionizing radiation responsible for reionization.  The
signal we predict, on the order of 
$l^2C_l/(2\pi)\sim 500-1000~{\rm \mu K}^2$, should be easily
detectable by correlating existing CMB maps, such as those produced by
the WMAP experiment, with maps produced by upcoming observations of the 21-cm
background with the Square Kilometer Array (SKA).  

Our derivation of the cross-correlation rests upon linear perturbation
theory and the reasonable
assumption that the sizes of ionized regions are much smaller than
scales corresponding to $l\sim 100$.  However, assuming that the sizes
of ionized regions are much smaller than the fluctuations responsible for
the signal we predict is not equivalent to assuming that the ionized
fraction is uniform.  Because our prediction depends on the
correlation between ionized fraction and density, $P_{x\delta}$, 
we have derived a simple approximate model for it (see
Appendix).  In future work, we will use large-scale simulations of
reionization to verify the accuracy of the relation we derive, and
perhaps to refine our analytical predictions.  Whatever the result of
more detailed future calculations, we are confident that the CMB
Doppler-21--cm correlation will open a new window into the high redshift
universe and shed light on the end of the cosmic dark ages.

\acknowledgments
We would like to thank Asantha Cooray and the anonymous referee for
useful comments.  E.~K. acknowledges support from an Alfred P. Sloan
Fellowship. M.~A.~A. is grateful for the support of a Department of
Energy Computational Science Graduate Fellowship. This work was
partially supported by NASA Astrophysical Theory Program grants
NAG5-10825 and NNG04G177G and Texas Advanced Research Program grant
3658-0624-1999 to P.~R.~S.

\appendix
\section{Density-ionization Cross-correlation}

The size of \ion{H}{2} regions during reionization is a function of the neutral
fraction: as the neutral fraction decreases, the typical size of
\ion{H}{2} regions increases, quickly approaching infinity as the
neutral fraction approaches zero and the \ion{H}{2} regions percolate.
As predicted by analytical and numerical studies of the large scale
topology of reionization (Furlanetto, Zaldarriaga, \& Hernquist 2005;
Iliev et al. 2005), the typical \ion{H}{2} region size approaches only
up to a few tens of comoving Mpc at even near 
the end of reionization.  Since we are interested in
epochs during which the ionized fraction is about a half, we can safely assume
for our purposes that the typical \ion{H}{2} region size is smaller
than the length scales of the fluctuations relevant here ($\sim 100$
Mpc).  In this case, the ionized fraction within a given volume can be
determined by considering only sources located inside that volume.  

Let us suppose that we take a region in the universe which has
an overdensity of $\delta$, where $\delta\ll 1$.
If we assume that each baryon within a collapsed object can ionize
$\zeta(\delta)$ baryons, then the ionized fraction within some volume can be
written as a function of its overdensity $\delta$,
\begin{equation}
\ln[1-x_e(\delta)]=-\zeta(\delta)f_{\rm coll}(\delta),
\label{xelocal}
\end{equation}
where $f_{\rm coll}(\delta)$ is a {\em local} fraction of 
the collapsed mass to mean mass density, which would be different
from the average collapsed fraction in the universe,
$f_{\rm coll}(0)$.
Note that this functional form correctly captures the behavior at low
and high ionizing photon to atom ratio.  For $\zeta f_{\rm coll}\ll
1$, $x_e\simeq \zeta f_{\rm coll}\ll 1$, which corresponds to all
the ionizing photons emitted within the volume ionizing atoms within
that volume, as expected before H II regions have percolated.  
For $\zeta f_{\rm coll}\gg 1$, which corresponds to many more ionizing
photons than atoms, $x_e\simeq 1$, as expected after percolation.
Given that we are only considering sources located within the region,
however, this expression is only an approximation during
percolation, when sources from outside of the volume become visible.
We emphasize that in any case equation (\ref{xelocal}) is based on a
simplifying assumption and does not capture many of the subtleties
included in more sophisticated models of reionization.

Here, we present two functional forms for $\zeta(\delta)$ which are
meant to bracket two important physical limits.  
The first limit we will refer to as the ``Str\"omgren limit'', while
the second we will refer to as the ``photon counting limit''.  In both
limits, we will assume that each hydrogen atom in a collapsed object
will produce $\epsilon_\gamma(z)$ ionizing photons.

\subsection{Str\"omgren Limit}
If we assume that a fraction $\eta_*(z)$ of collapsed gas is undergoing a
burst of star formation of duration $\Delta t_*(z)$, then the ionizing
photon luminosity per unit volume, $\dot{N}_\gamma$, is given by
\begin{equation}
\dot{N}_\gamma=\frac{\epsilon_\gamma \eta_*
{\overline{n}}(1+\delta)}{\Delta t_*}f_{\rm coll}(\delta),
\label{ngamma}
\end{equation}
where $\overline{n}$ is the mean density of the 
universe. The ``Str\"omgren limit'' is defined such that 
every recombination is balanced by an emitted photon; the following
equation therefore applies: 
\begin{equation}
\ln[1-x_e(\delta)]=-\frac{\dot{N}_{\gamma}}{\dot{N}_{\rm
rec}}=-\frac{\epsilon_\gamma \eta_*f_{\rm coll}(\delta)}
{\alpha c_l\Delta t_*{\overline{n}}(1+\delta)},
\label{recombination}
\end{equation}
where $\alpha$ is the recombination coefficient, $c_l$ is the clumping
factor, and $\dot{N}_{\rm rec}$ is the recombination rate per unit
volume in a {\em fully}-ionized IGM.  The last two terms in
equation (\ref{recombination}) are the ratio of photon luminosity
within a given volume to the number of recombinations per unit time
which would occur in that volume were it to be fully ionized.  
Note again that this expression ensures the proper behavior of $x_e$
in the low and high photon luminosity limits. Combining equations
(\ref{xelocal}) and (\ref{recombination}), we find 
\begin{equation}
\zeta(\delta)=\zeta_0(1+\delta)^{-1}\approx \zeta_0(1-\delta),
\end{equation}
where $\zeta_0\equiv \epsilon_\gamma \eta_*/(\alpha c_l\Delta
t_*{\overline{n}})$  and the approximation is valid in the limit
$\delta \ll 1$.  In 
deriving this relation, we have assumed that the clumping factor,
$c_l$, is independent of $\delta$.  
While it is unlikely that clumping
decreases with increasing $\delta$, it is plausible that it could
increase.  This would decrease the correlation between
density and ionized fraction, $P_{x\delta}$.  Because this term
typically dominates the cross-correlation (see \S~3.4), this
would have the effect of reducing the predicted signal.  

\subsection{Photon Counting Limit}
In the ``photon counting limit'', 
recombinations are not important in determining the
the extent of ionized regions.  Instead, it is the ratio of all
ionizing photons {\em ever} emitted to hydrogen atoms which determines
the ionized fraction.  The number density of photons that have been
emitted within a volume with overdensity $\delta$ is given by
\begin{equation}
n_\gamma(\delta) \equiv \epsilon_\gamma {\overline n}(1+\delta)f_{\rm coll}(\delta).
\label{counting}
\end{equation}
In the photon counting limit, we assume that the ionized fraction is
given by
\begin{equation} 
\ln\left[1-x_e(\delta)\right]=-\frac{n_\gamma(\delta)}{{\overline
n}(1+\delta)}=-\epsilon_\gamma f_{\rm coll}(\delta).
\end{equation}
Combining equations (\ref{xelocal}) and (\ref{counting}), we find
that $\zeta$ is independent of overdensity,
\begin{equation}
\zeta(\delta)=\zeta_0\equiv \epsilon_\gamma.
\end{equation}

\begin{figure}
\plotone{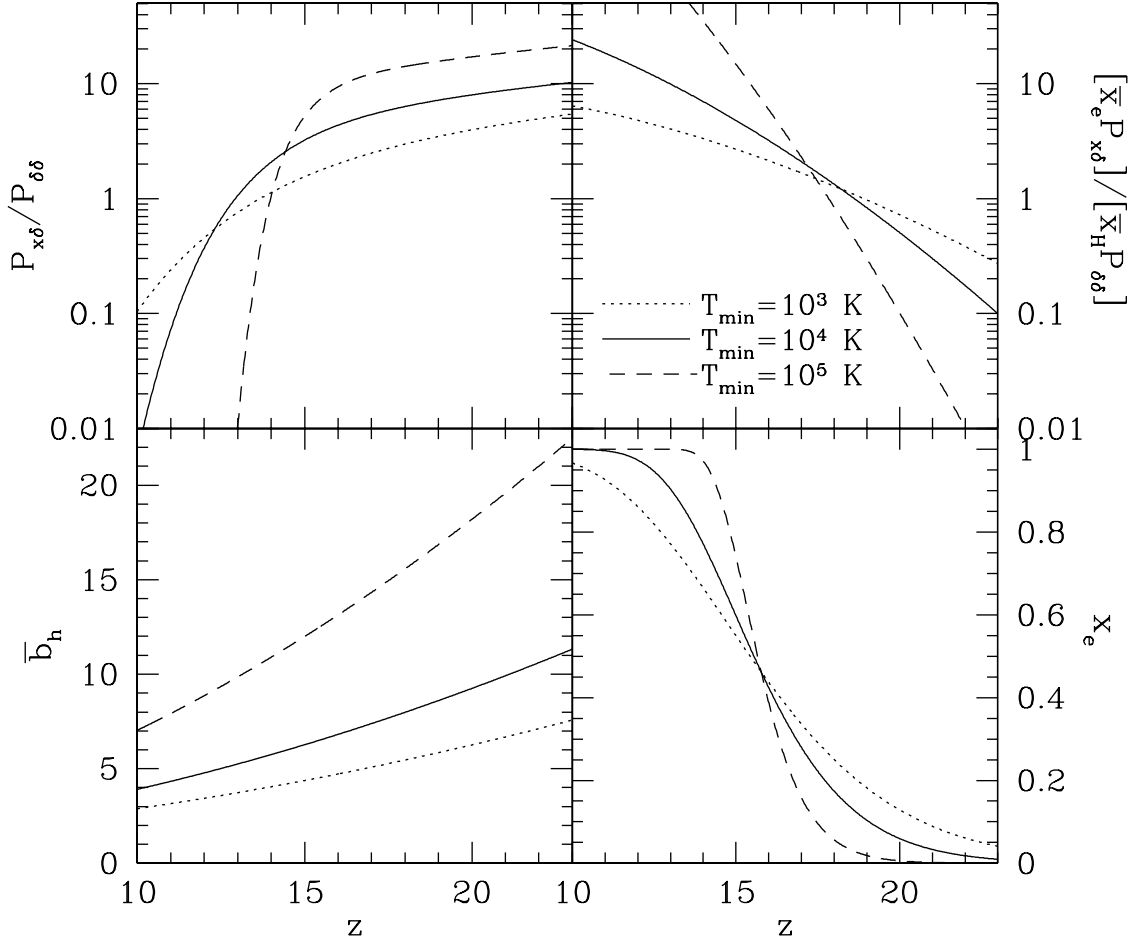}
\caption{The effect of bias on the relative importance of the
$P_{x\delta}$ and $P_{\delta\delta}$ terms of the cross-correlation
signal.  The dotted, solid, and dashed curves in each panel correspond
to models in which the sources responsible for reionization have a
minimum virial temperature of $10^3$, $10^4$, and $10^5$ K,
respectively.  ({\em bottom-right}) Shown are the reionization
histories with a value of $\zeta_0$ chosen such that the Thomson
scattering optical depth 
$\tau_{\rm es}=0.15$.   The model with $T_{\rm min}=10^4$ K is a
reionization history obtained with $\zeta_0=200$,  and is the same as
the single-reionization model presented in the main body of the paper.
({\em bottom-left}) Mean halo bias vs. redshift.  ({\em top-left})
Ratio $P_{x\delta}/P_{\delta\delta}$ as obtained using equation
(\ref{eq:cross_xd}).  ({\em top-right})  The ratio
$[{\overline{x}}_eP_{x\delta}]/[{\overline{x}}_HP_{\delta\delta}]$.  Note that
when the ratio equals one, the correlation vanishes, while larger
values indicate a dominant contribution from the $P_{x\delta}$ term.
The top panels assume the photon counting limit ($f=0$).
}
\smallskip
\label{biasplot}
\end{figure}

\subsection{Dependence of Collapsed Fraction on $\delta$}

Motivated by these two limits, we parameterize the $\delta$-dependence
of $\zeta$ as
\begin{equation}
\zeta(\delta)=\zeta_0(1-f\delta).
\label{zetadelta}
\end{equation}
In the photon counting limit, $f=0$, while in the recombination dominated
limit $f=1$.
Now that we have specified the form of $\zeta(\delta)$, we turn
to the collapsed fraction, $f_{\rm coll}(\delta)$.  The average
collapsed fraction in the universe, i.e., $f_{\rm coll}(\delta)$ 
with $\delta=0$, is given by 
\begin{equation}
f_{\rm coll}(0)={\rm erfc}\left[\frac{\delta_c(z)}
{\sqrt{2}\sigma_{\rm min}}\right].
\end{equation}
According to the extended Press-Schechter theory, the local collapsed
fraction is (Lacey \& Cole 1993)
\begin{equation}
f_{\rm coll}(\delta,m)={\rm erfc}\left[\frac{\delta_c(z)-\delta/D(z)}{\sqrt{2\left[\sigma^2_{\rm min}-\sigma^2(m)\right]}}\right],
\label{fcolll}
\end{equation}
where $m$ is the mass of the region.  
On large scales, $\sigma(m)\ll \sigma_{\rm min}$ and $\delta\ll 1$, so
that equation (\ref{fcolll}) can be expanded in a Taylor series around
$\delta=0$,
\begin{equation}
f_{\rm coll}(\delta)\simeq f_{\rm
coll}(0)+\sqrt{\frac{2}{\pi}}\frac{e^{-\delta_c^2(z)/2\sigma^2_{\rm
min}}}{\sigma_{\rm min}D(z)}\delta.
\label{fcolltaylor}
\end{equation}
An alternative expression for the local collapsed fraction can be
written in terms of the mean halo bias $\overline{b}_h$, 
\begin{equation}
f_{\rm coll}(\delta)=f_{\rm coll}(0)\frac{1+\overline{b}_h\delta}{1+\delta},
\end{equation}
which, for $\delta\ll 1$, is well-approximated by
\begin{equation}
f_{\rm coll}(\delta)\simeq f_{\rm coll}(0)
\left[1+(\overline{b}_h-1)\delta\right].
\label{fcollbias}
\end{equation}
Equations (\ref{fcolltaylor}) and (\ref{fcollbias}) are consistent
only if 
\begin{equation}
\overline{b}_h\equiv
1+\sqrt{\frac{2}{\pi}}\frac{e^{-\delta_c^2(z)/2\sigma^2_{\rm
min}}}{f_{\rm coll}(0)\sigma_{\rm min}D(z)}.
\label{bias}
\end{equation}
The reader can easily verify that this expression is the same as that
found by averaging over the halo bias derived by Mo \& White (1996), 
\begin{equation}
b(\nu)=1+\frac{\nu^2-1}{\delta_c},
\end{equation}
which gives
\begin{equation}
\overline{b}_h=\frac{\int_{\nu_{\rm min}}^\infty d\nu f(\nu)b(\nu)}
{\int_{\nu_{\rm min}}^\infty d\nu f(\nu)},
\end{equation}
where 
\begin{equation}
f(\nu)\propto \exp\left[-\nu^2/2\right].
\end{equation}
The Taylor expansion of the collapsed fraction used in deriving equation
(\ref{fcolltaylor}) is therefore fully consistent with the standard
linear bias formalism.

\subsection{Final Expression}
Equations (\ref{xelocal}), (\ref{zetadelta}), and (\ref{fcollbias}) imply that
\begin{equation}
\ln[1-x_e(\delta)]=\ln[1-\overline{x}_e]\left[1+(\overline{b}_h-1-f)\delta\right]
\end{equation}
where we have used the fact that $\ln[1-\overline{x}_e]=-\zeta_0
f_{\rm coll}(0)$. 
In the limit where $f=0$ and we are simply counting photons, the
ionized fraction does not depend on $\delta$ if the mean halo bias
${\overline{b}}_h=1$, since the additional photons emitted within a given
region are exactly canceled by the additional atoms contained within
that region.  If recombinations are important, however, the condition
for ${\overline{x}}_e$ to remain constant is given by
${\overline{b}}_h=2$.  In this case, the bias must compensate for the
additional photons necessary to balance the enhanced recombination
rate per atom within the volume.  As was noted in \S~A.1, this relies
on assuming a clumping factor which is independent of density.  If the
clumping factor is an increasing function of density, then the
condition would be ${\overline{b}}_h>2$.
The cross-correlation of ionized fraction and density fluctuations is
given by (for $\delta\ll 1$ and $\delta_x \ll 1$)
\begin{equation}
\langle \delta_x \delta\rangle \simeq
-\frac{1-\overline{x}_e}{\overline{x}_e}\ln(1-\overline{x}_e)
(\overline{b}_h-1-f)\langle \delta\delta\rangle,
\label{eq:dxd}
\end{equation}
so that
\begin{eqnarray}
\nonumber
{\overline x}_e P_{x\delta}(k)&\simeq&-(1-\overline{x}_e)
\ln(1-\overline{x}_e)(\overline{b}_h-1-f)P_{\delta\delta}(k)\\
&=&-\overline{x}_H
\ln \overline{x}_H(\overline{b}_h-1-f)P_{\delta\delta}(k).
\label{eq:cross_xd}
\end{eqnarray}
A comparison of the different terms implied by equation
(\ref{eq:cross_xd}) is shown in Figure \ref{biasplot}. 

If the universe
begins to recombine, then equation (\ref{eq:cross_xd}) would be
correct in the limit 
where the recombination time is short compared to the time it takes
for sources to dininish in intensity.  For simplicity, we will assume
this is the case and use the relation of equation (\ref{eq:cross_xd})
exclusively in the main body of the paper.  In the following section,
we will investigate the departure from that relation for the case
where the sources decay faster than the recombination time.

\subsection{Bias in a Recombining Universe}

Equation (\ref{eq:cross_xd}) was derived under the assumption that the
ionized fraction is determined by the abundance of reionization
sources and the density of their environment.  However, in the limit
where the intensity of ionizing radiation due to these sources drops
precipitously, as may be expected from metal enrichment or some other
form of negative
feedback, the ionized fraction will be determined by the rate of
recombination.  In order to understand the effect of a ``recombination
epoch'' on the cross-correlation, we will derive a simple relation for
$P_{x\delta}$ for the extreme case in which a region of the universe
recombines with no sources present.

In the absence of ionizing radiation, recombination is expected to
proceed according to
\begin{equation}
\frac{dx_e}{dy}=-(1+\delta)x_e^2,
\label{eq:recombine}
\end{equation}
where $y\equiv t/{\overline{t}}_{\rm rec}$ is time in units of the mean
recombination of the universe, ${\overline{t}}_{\rm rec}$.  We will
take the initial ionized fraction to be a deterministic function of the
overdensity, so that the initial fluctuation of ionized fraction (when
recombination begins occur) is $\delta_{x,i}=b_{x,i}\delta$,  where
the subscript $i$ refers to the initial value.  If the bias in ionized
fraction just before recombination begins is described by equation
(\ref{eq:dxd}), then we have 
\begin{equation}
b_{x,i}=-\frac{1-\overline{x}_{e,i}}{\overline{x}_{e,i}}\ln(1-\overline{x}_{e,i}) (\overline{b}_h-1-f).
\end{equation}
For the sake of generality, however, we will report our results in
terms of $b_{x,i}$.  Solving for equation (\ref{eq:recombine}), we
obtain the time evolution of $x_e$,
\begin{equation}
x_e(y)=\frac{x_{e,i}}{1+x_{e,i}(1+\delta)y}, 
\end{equation}
from which it follows that
\begin{equation}
\delta_x({\overline{x}}_e)=\left[\frac{{\overline{x}}_e}{{\overline{x}}_{e,i}}
\left(b_{x,i}+1\right)-1\right]\delta,
\end{equation}
where we have assumed $\delta\ll 1$ and have used the relation
\begin{equation}
{\overline{x}}_e(y)=\frac{{\overline{x}}_{e,i}}{1+{\overline{x}}_{e,i}y}.
\end{equation}
When ${\overline{x}}_e={\overline{x}}_{e,i}$, $\delta_x=b_{x,i}\delta$,
as expected.  As the universe recombines to become fully neutral,
${\overline{x}}_e\rightarrow 0$ and $\delta_x\rightarrow -\delta$.
Since we expect $b_{x,i}>1$ because overdense regions have an
overabundance of ionizing dources, a period of recombination is
expected to weaken the importance of $P_{x\delta}$ term.  When
${\overline{x}_e}/{\overline{x}}_{e,i}=1/2$ and $b_{x,i}\gg 1$, for
example, the bias determined from (\ref{eq:cross_xd}) is too large by
a factor of two.  Because we have assumed that sources turn off
instantaneously in equation (\ref{eq:recombine}), this is an upper
limit to the effect of a recombination epoch on $P_{x\delta}$. 

\end{document}